\documentclass[pra,twocolumn,english,floatfix]{revtex4-2}
\usepackage{blindtext}
\usepackage{amsmath}
\usepackage{amssymb}
\usepackage{graphicx}
\usepackage{amsfonts}
\usepackage{physics}
\usepackage{comment}
\usepackage{natbib}
\usepackage{color}
\usepackage{tikz}
\usepackage{amsthm}
\usepackage{dsfont}
\usepackage{cancel}

\begin{document}
\title{Comment on: ``Association between quantum paradoxes based on weak values
and a realistic interpretation of quantum measurements''}

\author{Juan Jose Seoane$^1$}
\author{Xabier Oianguren-Asua$^2$ }
\author{Albert Solé$^3$}
\author{Xavier Oriols$^1$}
\altaffiliation[email: ]{xavier.oriols@uab.es}
\affiliation{$^1$ Universitat Aut\`onoma de Barcelona, 08193 Bellaterra, Barcelona, Spain}
\affiliation{$^2$ Eberhard Karls Universität Tübingen, 72076 Tübingen, Germany }
\affiliation{$^3$  Universitat de Barcelona, 08001, Barcelona, Spain}

\begin{abstract}
In the paper [Phys. Rev. A 109, 022238 (2024)], Aredes and Saldanha analyze several paradoxes related to weak values and present a ``General Argument" that aims to show that ``realistic interpretations [...] of weak values lead to inconsistencies."  Although we agree with the identified inconsistencies for the specific weak values analyzed there, in this comment, we demonstrate that the origin of these inconsistencies is not their ``General Argument," which is formally incorrect. We use Bohmian mechanics as a counterexample to confirm that their conclusions are not valid for all weak values and quantum theories. In particular, we show that weak values post-selected in position can, in fact, be interpreted within Bohmian mechanics as properties of quantum systems, detached from any measuring devices, in a consistent and meaningful way.
\end{abstract}

\maketitle
\section{Introduction}

The paper in Ref. \cite{association2014pra}  initiates a discussion on whether weak values can be interpreted as referring to real properties within contextual quantum theories. The authors explore the meaning of weak values by analyzing several paradoxes. While we agree with Aredes and Saldanha that a realistic interpretation of weak values, as applied to the specific paradoxes they examine, leads to inconsistencies, we disagree with their explanation for the origin of these inconsistencies, as presented in the crucial section entitled ``General Argument" of Ref. \cite{association2014pra}.

The authors define a ``\textit{realistic interpretation of a quantum measurement}" (RIQM) as follows:
``\textit{A measurement performed on a quantum system reveals the underlying ontological value of the measured quantity, that continues the same after the measurement}''. In addition, they define  a ``\textit{realistic interpretation of weak values}" (RIWV) as follows: ``\textit{The weak value $\langle O \rangle_w$ of an operator $O$ reveals the objective reality of the physical quantity $\tilde O$ associated to this operator}'' \cite{footnote0b}.  From these definitions, through the above-mentioned ``General Argument," they conclude that:
\begin{itemize}
\item \textbf{Conclusion 1}: ``\textit{The attribution of a physical reality to the weak value $\langle O \rangle_w$ is equivalent to adopting the cited realistic interpretation of quantum measurement}''. In other words, they consider that RIQM and RIWV are equivalent when dealing with weak values. 
\end{itemize}
Aredes and Saldanha consider that RIQM is ``\textit{highly controversial}'' since it plainly contradicts the well-known quantum contextuality. Now, given this fact and the alleged equivalence between RIQM and RIWV, the authors finally conclude that: 
\begin{itemize}
\item \textbf{Conclusion 2}: ``\textit {realistic interpretations [...] of weak values lead to inconsistencies}". 
\end{itemize}

In this comment, we present three results to show that both \textbf{Conclusion  1} and \textbf{Conclusion 2} are incorrect. 

\section{Fallacious argumentation}
\label{s2}

Let us carefully reconstruct Aredes and Saldanha's argument in Sect. II (entitled ``General Argument'') of Ref. \cite{association2014pra} to show that it is fallacious. The authors invite us to consider a weak measurement procedure of an observable $O$, in which the quantum system is pre-selected in a state $|\psi_i\rangle$ and post-selected in a state $|\psi_f\rangle$. The obtained weak value is therefore: $\langle O \rangle_w = {\langle \psi_f | O | \psi_i \rangle}/{\langle \psi_f | \psi_i \rangle}$. Next, the authors consider a situation in which the operator $\hat{O}$ can be written as the sum of an operator $\hat{P}$ from which the pre-selected state is an eigenvector with eigenvalue $p$, and an operator $\hat{Q}$ from which the post-selected state is an eigenvector with eigenvalue $q$. That is, $\hat{O}$ satisfies $\hat{O} = \hat{P} + \hat{Q}$  with $\hat{P}|\psi_i\rangle = p|\psi_i\rangle$  and $\hat{Q}|\psi_f\rangle = q|\psi_f\rangle$ (see their Eq. (2) in Ref. \cite{association2014pra}). As a piece of terminology, the authors use the symbol $\tilde {O}$ to denote the ``ontological value'' corresponding to the property $O$ associated with the operator $\hat {O}$. These ontological values are interpreted as objective properties of the considered system and are mentioned both in RIQM and in RIWV. Now, we are prepared to present the crucial steps of Aredes and Saldanha's argument.

First, assuming RIWV, they claim that, in the scenario above, the objective physical quantities associated with the operators $\hat{O}$, $\hat{P}$ and $\hat{Q}$ are \textit{$\tilde O=p+q$}, \textit{$\tilde P=p$} and \textit{$\tilde Q=q$} respectively. Second, assuming RIQM, and through a slightly more elaborate reasoning, the authors claim to reach the same conclusion, namely, that \textit{$\tilde O=p+q$}, \textit{$\tilde P=p$} and \textit{$\tilde Q=q$}. So, up to this point, we can see that the authors' claim is that both RIWV and RIQM entail the same consequence with respect to the ontological values $\tilde{O}$, $\tilde{P}$ and $\tilde{Q}$ in the considered scenario. Rather surprisingly, from this, they conclude:

\begin{quote}
``\textit{The attribution of a physical reality to the weak value value } [what we named RIWV] \textit{is equivalent \cite{footnote11} to adopting the cited realistic interpretation of quantum measurements} [what we referred to as RIQM] \textit{since both assumptions lead to the same objective value for the physical quantity} $\tilde O:\tilde O = p + q$ [...] \textit{Since the cited realistic interpretation of quantum measurements} [RIQM] \textit{is highly controversial, a reasonable way to avoid all the cited quantum paradoxes is to deny this realistic interpretation of quantum measurements} [RIQM]\textit{,  also denying the realistic view of the weak values} [RIWV].''
\end{quote}

Now, it should be clear that this reasoning is fallacious. Two assumptions can share some consequences, yet this does not entail that they are logically equivalent. Given that Aredes and Saldanha reach \textbf{Conclusion 1} in a fallacious manner, this conclusion is unfounded. In the next session we demonstrate that, moreover, it is false.\vspace{-0.4cm}

\section{Conclusion 1 is incorrect}\vspace{-0.2cm}
\label{s3}

Bohmian mechanics is a theory that accounts for all non-relativistic quantum phenomena. Its ontology consists of particles with well-defined positions regardless of whether or not they are being observed \cite{bohm1952a,bohm1952b,chapter1,chapter2}. The dynamics involve a (deterministic) equation of motion for the particles' positions (the so-called "guiding equation" \cite{chapter1}) which provides the velocity of each particle as a functional of the wave function. Additionally, the law of motion for the wave function itself is the (deterministic) Schrödinger equation.

\subsection{Bohmian mechanics, a contextual theory}
\label{s31}

When Aredes and Saldanha mention that RIQM is ``highly controversial,'' they explicitly refer to quantum contextuality and the Kochen-Specker theorem \cite{mermin1993} to support this claim. One standard way to avoid the refutatory charge of this theorem is to accept that any hidden-variable theory aiming to be empirically adequate must be \textit{contextual}. This amounts to denying that, if a quantum system possesses a property (the ontological value of an observable), it does so independently of any measurement context. It is worth noticing that Bohmian mechanics is contextual in this sense and that, as a consequence, is not refuted by the Kochen-Specker theorem and other similar results \cite{hardy1996}.

RIQM affirms (or entails) what is known in the literature as the ``\textit{Faithful Measurement Principle}'', which asserts that \textit{a measurement of an observable faithfully reveals the value which that observable had immediately prior to the measurement interaction} \cite{held2022}. Quantum mechanics violates both the ``\textit{Faithful Measurement Principle}'' and RIQM. In order to show this, consider a system in the state $\frac{1}{\sqrt{2}} \left( |\psi_a\rangle + |\psi_b\rangle \right)$,
where $\hat{O}|\psi_a\rangle = a|\psi_a\rangle$ and $\hat{O}|\psi_b\rangle = b|\psi_b\rangle$. Now, suppose one makes a (strong) measurement of observable $O$ and obtains the eigenvalue $a$ as the outcome. According to orthodox quantum mechanics and the eigenvalue-eigenstate link, this measurement does not faithfully reveal the pre-existing value of $O$ because, prior to the measurement, the ontological value of $O$ was simply undefined. Therefore, in orthodox quantum mechanics, it is the very act of measuring $O$ that ``creates'' a definite ontological value of the property $O$.

Bohmian mechanics also violates both the  ``\textit{Faithful Measurement Principle}'' and RIQM. In this theory, the ontological value of a property after a measurement typically does not match its pre-existing ontological value because the value of the property is modified during the interaction with the measuring apparatus. Notice that it is \textit{not }the case that a definite property is ``created" by the very act of measuring, as happens according to the orthodox quantum-mechanical approach. Simply, the (measuring) interaction between the two systems may eventually lead to a change in some of their properties.

With all, RIQM is not only ``highly controversial'' but also incompatible with any quantum theory, as it contradicts the inherently perturbative nature of quantum measurements (and quantum interactions in general). This characteristic reveals a form of contextuality broader than that typically discussed within the framework of the Kochen-Specker theorem. Throughout the rest of our paper, we use the term `contextuality' in this broader sense. 

\subsection{Bohmian mechanics as a counterexample to Conclusion 1}
\label{s33}

We now consider an operator $\hat{O}=\hat{P}+\hat{Q}$ that satisfies the pre-requisites of the General Argument in Ref. \cite{association2014pra} and whose weak values are physically meaningful within the framework of Bohmian mechanics.

As the operator $\hat{Q}$ we take the position operator $\hat X$ so that we consider for post-selection the position eigenstate $|x_o\rangle$. We have that \cite{footnote8}:
\begin{equation}
\langle X \rangle_w=\Re{\frac{\langle x_o|\hat X|\psi_i(t)\rangle}{\langle x_o|\psi_i(t)\rangle}}=x_o,
\label{pos}
\end{equation}
where $\hat X|x_o\rangle =x_o |x_o\rangle$. As operator $\hat{P}$ we take the velocity operator, $\hat V=-i\frac{\hbar}{m}\frac{\partial}{\partial x}$, with $m$ being the mass of the particle. For the post-selected position eigenstate  $|\psi_f\rangle=|x_o\rangle$, one obtains the weak value $\langle V \rangle_w$ given by:
\begin{eqnarray}\hspace{-0.6cm}
 \langle V \rangle_w=\Re{\frac{\langle x_o|\hat V|\psi_i(t)\rangle}{\langle x_o|\psi_i(t)\rangle}}=\frac{J(x_o,t)}{|\psi_i(x_o,t)|^2}=v^{\psi_i}(x_o,t)
\label{velo}
\end{eqnarray}
where $J(x,t)=\frac{\hbar}{m}\Im{\psi_i(x,t)\frac{\partial \psi^*_i(x,t)}{\partial x}}$ is the quantum expression of the current density \cite{wiseman2007}. Within Bohmian mechanics, it is \textit{natural} to interpret realistically  $x_o$ in \eqref{pos} and $v^{\psi_i}(x_o,t)$ in \eqref{velo} as the position and the velocity, respectively, of a particle described by the wave function  $\psi_i(x,t)=\langle x|\psi_i(t)\rangle$ and located at $x_0$.

In the case of a plane wave $\langle x|\psi_i\rangle \propto e^{ikx}$ (or an approximate plane wave), one gets $v^{\psi_i}(x_o,t)=\hbar k/m$. Then, one can construct the operator $\hat O=\hat V+\hat X$ which satisfies the requirements of Eqs. (2) and (3) in the ``General Argument''  of Ref.  \cite{association2014pra} for the given wavefunction. Within Bohmian mechanics, we obtain  $\langle O\rangle_w =\hbar k/m+x_o$ and this weak value can be readily interpreted as the sum of the real velocity and the real position of the particle, satisfying RIWV. And still, thanks to the contextual nature of the Bohmian theory, it does not satisfy Aredes and Saldanha's RIQM (i.e. RIWV is possible without RIQM), serving as a straightforward counterexample demonstrating that \textbf{Conclusion 1} cannot be correct.

\section{Conclusion 2 is incorrect}
\label{s4}


Orthodox and Bohmian quantum mechanics are empirically equivalent. Born's rule follows as a corollary from the equivariant motion of Bohmian trajectories and the quantum equilibrium hypothesis \cite{durr1992}. In particular, for a set of hypothetical trajectories $x^j(t)$, with $j\in\{1,...,N\}$, each representing the trajectory of the system in a different experimental realization of the same single-particle wave function $\psi(x,t)$, one obtains that:
\begin{equation}
|\psi(x,t)|^2 = \lim_{N\to\infty} \frac{1}{N}\sum_{j=1}^N \delta(x-x^j(t)).
\label{modal0}
\end {equation}

Apart from the position of a particle, it is \textit{natural} to define additional properties subordinate to this position, as we have done for the velocity $v^{\psi}(x,t)|_{x=x^j(t)}={dx^j(t)}/{dt}$ in Section \ref{s33}. In general, for any function $S^\psi_B(x,t)$ of the wavefunction, one can define $S_B^\psi(x^j(t),t)$ as a Bohmian property subordinate to $x^j(t)$ (be it ontologically meaningful or merely informative). As happens to $x^j(t)$ and $v^{\psi}(x^j(t),t)$, this new Bohmian property refers to quantum systems that are not being measured. And still, within the Bohmian theory, using \eqref{modal0}, one can compute the ensemble value of the property  $S_B^\psi(x^j(t),t)$ over all $x^j(t)$ to be:
\begin{equation}
\langle {S}_B \rangle= \lim_{N\rightarrow\infty} \frac{1}{N} \sum_{i=1}^N S_B^\psi(x^j(t),t) = \int dx \;S_B^\psi(x,t)|\psi(x,t)|^2.
\label{modal2} 
\end{equation}
Among the possible fields $S^\psi_B(x,t)$, there are some, the so-called {\em local expectation values} of Hermitian operators \cite{Holland}, which yield physically relevant properties for the trajectories because their ensemble average gives exactly the expectation associated to quantum measurements of the operator in question. Precisely, the local expectation value of the Hermitian operator $\hat{S}$ is defined to be:
\begin{eqnarray}
    S^\psi_B(x,t)&:=&\Re{\frac{\langle x|\hat{S}|  \psi(t) \rangle}{\langle x|  \psi(t) \rangle}}
    \label{modal3}
\end{eqnarray}
which is straightforwardly seen to satisfy $\langle S_B^\psi \rangle=\langle \psi |\hat{S}|\psi\rangle$. Yet, \eqref{modal3} is nothing but the position post-selected weak value of $\hat{S}$ \cite{footnote8}. It turns out that not only for the position and velocity, but for many more observables, their position post-selected weak values, i.e., local expectations \eqref{modal3}, not only match the operator expectations, but happen to coincide with the Bohmian definitions of those properties, which is found independently of the discussion of a measurement of weak values (with no inconsistencies whatsoever). For instance, if one sets $\hat{S}$ to be the Hamiltonian operator $H$, the local expectation turns out to be exactly equal to the Bohmian energy (kinetic plus classical and quantum potentials \cite{chapter1}) of the particle. For the angular momentum operator, one gets the angular momentum of the Bohmian particle. And the list goes on \cite{devashish2021,chapter1, Holland}.  

In summary, interpreting weak values post-selected in position as values of properties of Bohmian particles offers a consistent and paradox-free framework for understanding experiments on the weak values defined as $S_{B}^\psi(x,t)=\Re{{\langle x|\hat{S}|  \psi(t) \rangle}/{\langle x|  \psi(t) \rangle}}$. This consistency is directly inherited from the consistency of the Bohmian theory in accounting for all non-relativistic quantum phenomena \cite{chapter1,chapter2,footnote4}. Therefore, we have shown that \textbf{Conclusion 2} is also incorrect as a general claim. At least, a realistic interpretation of weak values post-selected in position within Bohmian mechanics does not lead to any contradiction \cite{footnote20}.

\section{``Non-contextual weak values" from contextual Bohmian mechanics}
\label{s5}

We have claimed that weak values (post-selected in position), as formal quantities (independently of  experimental protocols estimating them), can be consistently interpreted within Bohmian mechanics as properties determinately possessed by particles --such as position, velocity etc. Now, as explained later, there exist experimental protocols to estimate some of these weak values for microscopic systems. So, some of their Bohmian properties, which are a priori independent of any context in the laboratory, can be determined empirically. If we call this a ``measurement", it will seem there is a contradiction with the fact that Bohmian mechanics is a contextual quantum theory. But there is no contradiction when we clarify that the word ``measurement" can have slightly different meanings. In quantum mechanics, it usually means an experiment conducted on a single copy of the system. But, a protocol to determine a weak value uses an ensemble of copies of the system, with a contextual intervention on each of them. Our intervention will alter the {\em subsequent} dynamics of each copy for the inherent backaction of quantum interactions, but within Bohmian mechanics, the obtained number (the weak value) will indeed coincide with each copy's property at the time right {\em before} our intervention started. The weak value is ``measured" in this sense. 



To be specific, the empirical protocol to ``measure" a weak value involves an ensemble of $N$ identically prepared microscopic systems, all described by the same initial wave function $\psi$. On each individual microscopic system, a weak measurement \cite{footnote7} of $\hat{O}$, giving $o_a$, is performed, followed by a projective measurement of an $\hat{F}$ giving $f_a$ (with eigenvector $|f
 _a\rangle$). This yields a large set of pairs $\{o_a, f_a\}_{a=1}^N$. The average of the $o_a$'s conditioned on a fixed $f_a=f$ then estimates $\Re{\langle f|\hat{O} \psi \rangle/\langle f|\psi\rangle}$ (to leading order in both the device-subsystem coupling and the time between each $o_a$ and $f_a$ measurement) \cite{footnote8, chapter2,chapter1}. 

In particular, the number estimated by considering $\hat{O}=\hat{v}$ and $\hat{f}=\hat{X}$ is  $v^\psi(x,t)$ (as in equation \eqref{velo}), i.e., the Bohmian velocity that a microscopic system with wavefunction $\psi(x,t)$ (free from the perturbation of a measurement back-action) would have. Note that even if this will approximate the actual Bohmian velocity of the individual microscopic systems used to compute $v^\psi(x,t)$ {\em right before the measurements started} in the laboratory, typically, it does not allow the determination of their Bohmian velocity {\em between the two measurements nor at later times.} A quantum measurement (weak or not), implies an entanglement between each system and measurement apparatus. As soon as this interaction begins, the  Bohmian velocity of each microscopic system needs to be considered in each joint system-measurement apparatus configuration space (namely, each system's velocity will no longer be only a function of $x$, as in $v^\psi(x,t)$, but will also depend on external variables in the laboratory).  Note that none of these should be surprising, since the experimental protocol yielding the theoretical expectation value $\langle \psi|O|\psi\rangle$ also satisfies: (i) the outcome is obtained by averaging, (ii) it characterizes how the measured copies of the microscopic system were right before they were intervened, but (iii), not (usually) how they were after that.

 \section{Acknowledgements}
J.J.S. is supported by Grant No. PRE2022-104030, funded by MCIN/AEI/10.13039/501100011033 and FSE+. X.O acknowledges support from the Spain’s Ministerio de Ciencia, Innovación y Universidades under Grants PID2021-127840NB-I00 (MICINN/AEI/FEDER, UE) and PDC2023-145807-I00 (MICINN/AEI/FEDER, UE), and European Union’s Horizon 2020 research and innovation program under Grant 881603 GrapheneCore3. The research conducted by A.S. is funded by Grant PID2020-115114GB-I00 and Grant CEX2021-001169-M from MCIN/AEI/10.13039/501100011033.

\nocite{*}

\bibliographystyle{apsrev4-1} 

\begin{thebibliography}{99}
\bibitem{footnote8} The expression ${\langle x|\hat S|  \psi(t) \rangle}/{\langle x|  \psi(t) \rangle}$ inside Eq. \eqref{modal3} is a complex number and it can be written as $\Re{{\langle x|\hat S| \psi(t) \rangle}/{\langle x| \psi(t) \rangle}}+i\Im{{\langle x|\hat S| \psi(t) \rangle}/{\langle x|  \psi(t) \rangle}}$. Thus, one can define one weak value for the real part of  \eqref{modal3} and another weak value for the imaginary one. The experimental determination of these two weak values involves two different protocols in the laboratory. In this comment, we only deal with the weak value associated with the real part of \eqref{modal3}. But a similar discussion, if needed, can be done for the imaginary part \cite{chapter2}. Notice, in addition, that similar conclusions can be obtained for weak values pre-selected (instead of post-selected) in position defined as $\Re{{\langle \psi(t)|\hat S| x \rangle}/{\langle \psi(t)| x \rangle}}$.
\bibitem{footnote7} The weak measurement of $\hat{O}$ is described by a Positive Operator-Valued Measure given by a projective measurement on an ancillary system weakly coupled to the object quantum system. The measurement collapses the ancilla into an eigenstate of the observable coupled to the operator $\hat{O}$ of the subsystem, yielding the eigenvalue $o_a$ of the former as the outcome of the experiment.
\end{thebibliography}

\end{document}